\documentclass[sn-mathphys-num]{sn-jnl}

\usepackage{graphicx}%
\usepackage{multirow}%
\usepackage{amsmath,amssymb,amsfonts}%
\usepackage{amsthm}%
\usepackage{mathrsfs}%
\usepackage[title]{appendix}%
\usepackage{xcolor}%
\usepackage{textcomp}%
\usepackage{manyfoot}%
\usepackage{booktabs}%
\usepackage{algorithm}%
\usepackage{algorithmicx}%
\usepackage{algpseudocode}%
\usepackage{listings}%
\usepackage{tikz}
\usepackage{pgfplots}

\usepackage{dirtytalk}

\theoremstyle{thmstyleone}%
\theoremstyle{thmstyletwo}%
\newtheorem{remark}{Remark}%

\theoremstyle{thmstylethree}%
\newcommand{\del}{\partial}

\newcommand{\dif}{~\mathrm{d}}

\begin{document}

\title[Semi-Analytic Solutions to the Noh Problem with a Black Box Equation of State]{Semi-Analytic Solutions to the Noh Problem with a Black Box Equation of State}


\author[1]{\fnm{Seth} \sur{Gerberding}}\email{gerberding@lanl.gov}
\author[2]{\fnm{Jeff} \sur{Peterson}\email{jhp@lanl.gov}}
\author[1]{\fnm{Jim} \sur{Ferguson}\email{jmferguson@lanl.gov}}
\author[3]{\fnm{Scott} \sur{Ramsey}\email{ramsey@lanl.gov}}



\affil[1]{\orgdiv{XCP-8}, \orgname{Los Alamos National Laboratory}, \orgaddress{\city{Los Alamos}, \postcode{87544}, \state{New Mexico}, \country{U.S.A}}}
\affil[2]{\orgdiv{XCP-2}, \orgname{Los Alamos National Laboratory}, \orgaddress{\city{Los Alamos}, \postcode{87544}, \state{New Mexico}, \country{U.S.A}}}
\affil[3]{\orgdiv{XTD-NTA}, \orgname{Los Alamos National Laboratory}, \orgaddress{\city{Los Alamos}, \postcode{87544}, \state{New Mexico}, \country{U.S.A}}}





\abstract{The objective of this paper is to derive a method of constructing semi-analytic solutions to the Noh Problem when the equation of state (EoS) is a black box. Such solutions can be used for verification tests of hydrodynamics codes. We present the underlying theory, the method for finding solutions, and several examples of derived semi-analytic solutions. We end by performing a classic convergence test using a non-trivial semi-analytic solution.
}




\maketitle

\section{Introduction}\label{sec: introduction}

The objective of this paper is to present and demonstrate a method for deriving semi-analytic solutions to the Noh Problem that is agnostic to the geometry, initial conditions, and equation of state. In particular, we detail a method that only assumes a ``black-box'' equation of state (EoS) to solve the Rankine-Hugoniot jump conditions for the shocked variables and shock speed. The semi-analytic solution is an immediate consequence; these solutions can be used to perform verification of hydrodynamics codes. 

\par Verification, in a broad sense, is the process of ensuring a computational model faithfully reproduces solutions to a mathematical model of the system. This process generally consists of two tasks \cite{Oberkampf_Trucano_Hirsch_04}. The first consists of removing programming defects, or \textit{code verification}. The second consists of quantifying numerical errors, or \textit{solution verification} \cite{Roy_05}. Solution verification can take several forms, including comparing to exact solutions to the partial differential equation (PDE) and the method of manufactured solutions; see Freno et. al. \cite{Freno_21}. In this work, we focus on exact solution analysis. This analysis consists of comparing the numerical solution of the computational model to an exact solution of the mathematical model. This approach, while powerful, is somewhat limited: exact solutions are hard to come by. 

\par A more general category of solutions, called \textit{semi-analytic solutions}, can also be used. Semi-analytic solutions are \say{semi} insofar as they are an approximation of the exact solution, but that approximation is \textit{not} from a numerical discretization of the governing PDE. Consider solving a problem, \(G(x) = 0\). Suppose an implicit solution is found, meaning there is a relation \(x = f(x)\). To find \(x\) explicitly, a solver is required, but since the solver is often iterative, the solution is found within some tolerance. In our case, our solution is semi-analytic insofar that we have to solve a nonlinear system of algebraic equations for a set of coefficients. 

\par Our governing model is the compressible Euler equations. Finding exact solutions to these equations is notoriously difficult---existence and uniqueness are not always known---but a few exist, including the Sedov blast problem \cite{sedov_18}, the Guderley problem \cite{Ramsey_Kamm_12}, the Coggeshall problems \cite{Coggeshall_91}, and the Noh problem \cite{NOH1987}. 

\par One complicating factor for the Euler equations is the equation of state, which is a mathematical object that encodes various material and thermodynamical properties. A different equation of state means a different solution. From a verification perspective, it is desirable to perform verification tests as close to practical settings as possible. This problem is challenging since many materials do not have an equation of state in a simple closed form. Rather, they are algebraically complicated or even tabulated.

\par Traditional solutions to the Euler equations use the ideal gas equation of state, though significant work has been devoted to deriving exact solutions when the EoS is different. Particular attention is paid to the Noh problem. In one of the keystone papers, Axford leverages Lie Group theory to derive exact solutions to the Noh problem based on the ideal gas, stiffened gas, and Mie-Gruneisen equations of state \cite{axford_2000}. Ramsey et. al. expand this approach to two-parameter EoSs (such as Noble Able) and another form of Mie-Gruneisen \cite{Ramsey2017} EoSs. Additionally, they discover important constraints on the EoS for a solution to exist in cylindrical and spherical geometries. Later, Burnett, Honnell, and Ramsey analyze the Caranhan-Starling equation of state and derive a semi-analytic solution \cite{Burnett_17}. We note that these solutions were derived under various assumptions on the EoS parameters, for example the reference density in the stiffened gas EoS. 

\par This paper is a continuation of that overall project: to derive solutions to the Noh problem. We expand previous work by assuming a \say{black box} equation of state. That is, we only assume we have access to specific internal energy as a function of density and pressure (\(e = e(\rho, P)\)) or pressure as a function of energy and density (\( P = P(\rho, e)\)). Such assumptions are not new in terms of theory: Clayton et. al. develop a numerical algorithm that preserves invariant domains of the Euler equations but does so with a black box equation of state \cite{Clayton_Et_al_2023}. However, the authors were only able to perform verification tests using the van der Waals equation of state because there was a lack of solutions to test against; it is exactly this gap we aim to fill. Using black box EoSs expands abilities to produce semi-analytic solutions by removing assumptions on the form of the EoS, which can range from simple algebraic relations to tables of data, and EoS parameters, such as the reference density or reference pressure.

\par The rest of this paper is organized as follows. Section \ref{sec: theory} contains the theory that underpins our technique. Section \ref{sec: resultgs} presents semi-analytic solutions derived using the method, in particular for new equations of state such as the Steinberg/Mie-Gruneisen EoS and the \verb|SESAME| database. We also present results of a verification test performed using a semi-analytic solution. Section \ref{sec: conclusion} discusses conclusions and future work.

\section{Theory} \label{sec: theory}

\subsection{The Noh problem} \label{sec: noh_problem}
The Noh problem is a classic hydrodynamics problem for which analytic solutions exist. The problem consists of a fluid mass with constant density, pressure, and velocity, colliding with a symmetry point (a planar wall in Cartesian coordinates, an axis in cylindrical coordinates, and a point in spherical coordinates). At time \(t>0\), a shock forms and emanates from the symmetry point into the still oncoming fluid. The shock travels with constant speed and leaves behind a fluid state with zero velocity and constant density and constant pressure. In this work, we are concerned with the ``classic'' Noh problem, meaning the solution consists of two regions separated by a shock, the shock travels with constant speed, the shocked region has zero velocity, and the unshocked region retains the initial velocity. We refer to solutions with this structure as ``Noh solutions." Therefore, when we say a solution does not exist, or a particular problem set-up does not admit a solution, we do not mean that a solution of \textit{any} form does not exist, only that a solution of the aforementioned structure does not exist. There is work that extends solutions to the Noh problem when different assumptions are used. For example, Velikovich et. al show that self-similar solutions exist for the Noh problem when the initial velocity is not uniform and the fluid is dissipationless. The resulting solution in the unshocked region has a velocity less than the initial velocity and the shocked variables follow a different formula \cite{Velikovich_18}. 

\par The problem stems from the one-dimensional symmetric form of the compressible Euler equations, given by 

\begin{subequations}
    \begin{equation}
        \frac{\del \rho}{ \del t} + u \frac{\del \rho}{ \del x} + \rho \left(\frac{\del u}{ \del x}  + \frac{ m u}{x}\right) = 0, \label{eqn: euler-density} 
    \end{equation}
    \begin{equation}
        \frac{\del u}{\del t} + u \frac{\del u }{ \del x} + \frac{1}{\rho} \frac{ \del P}{\del x} = 0, \label{eqn: euler-u}
    \end{equation}
    \begin{equation}
        \frac{\del P}{\del t} + u \frac{ \del P}{\del x} + K \left( \frac{ \del u}{\del x} + \frac{m u}{x}\right) = 0, \label{eqn: euler-P}
    \end{equation}
\end{subequations}

\noindent where \(\rho\) is the density, \(P\) the pressure, \(u\) the velocity, \(e\) the specific internal energy, \(K\) the adiabatic bulk modulus, and \(m\) the space index (0 for planar, 1 for cylindrical, and 2 for spherical). We let \(x\) denote the radial coordinate. Following the above problem description, we have \(\rho_0 = \text{const}\), \(P_0 = \text{const}\), and \( u_0  = \text{const} < 0\). We let \(\rho_L, P_L, e_L\) denote the shocked density, pressure, and specific internal energy respectively; \(\rho_R, P_R, e_R\) denote their unshocked counterparts. Finally, we denote the shock speed by \(D\). 

\subsection{General Solutions} \label{subsec: general-solution}
The Noh problem is solved by leveraging Lie Groups and symmetries to pose the problem in terms of a self-similar variable: 

\begin{equation}
 \xi := \frac{a x}{t}. \label{def: self-similar-variable}    
\end{equation} 

\noindent Using this variable, the PDE can be posed a system of ODEs: 

\begin{subequations}
    \begin{equation}
        -\xi \frac{\dif \rho }{\dif \xi} + a u \frac{\dif \rho}{\dif \xi} + \rho \left( a \frac{ \dif u }{\dif \xi} + \frac{m a u}{\xi}\right) = 0, \label{eqn: euler-ode-rho}
    \end{equation}
    \begin{equation}
        - \xi \frac{\dif u}{\dif \xi} + au \frac{\dif u}{\dif \xi} + \frac{a}{\rho} \frac{\dif P}{\dif \xi} = 0, \label{eqn: euler-ode-u}
    \end{equation}
    \begin{equation}
        - \xi \frac{\dif P}{\dif \xi } + a u \frac{\dif P}{\dif \xi} + K \left( a \frac{ \dif u }{\dif \xi } + \frac{mau}{\xi}\right) =0. \label{eqn: euler-ode-p}
    \end{equation}
\end{subequations}

\par Following the work of Burnett et. al. and Ramsey et. al, the solution to the Noh problem then consists of solving the system of ODEs in the shocked and unshocked region \cite{Burnett_17, Ramsey2017}. In the unshocked region, the solution is given by: 

\begin{subequations}
    \begin{equation}
        \rho_R(x,t) = \rho_0 \left(1 - \frac{u_0 t}{x} \right)^m ,\label{eqn: unshocked-rho} 
    \end{equation}
    \begin{equation}
        P_R(x,t) = P_0, \label{eqn: unshocked-P}
    \end{equation}
    \begin{equation}
        u_R = u_0.\label{eqn: unshocked-u}
    \end{equation}
\end{subequations}

\par All of the shocked variables are constant: \(\rho_L = \text{const}\), \(P_L = \text{const}\), \(e_L = \text{const}\), and \(u_L = 0\). The geometry specific solutions are presented below. 

\par Before we proceed, we must recall an important theoretical constraint on the Noh problem. In particular, there is a restriction on the initial conditions for a Noh solution to exist in cylindrical and spherical geometries \cite{Burnett_17}. The restriction is the following: when \( m \neq 0\), then \(P_0=0\) must be the initial pressure \textit{and} \(e(0, \rho) \equiv \text{const}\). When \( m=0 \), there are no restrictions.

\subsubsection{General Solution in Planar Geometry}

As there are no restrictions on the initial conditions and EoS, the general solution to the Noh problem in planar geometry is given by: 

\begin{subequations}
    \begin{equation}
        \rho(t,x) = \begin{cases}
            \rho_L & \frac{x}{t}< D \\
           \rho_0  & \frac{x}{t} > D
        \end{cases}, \label{eqn: planar-density-solution}
    \end{equation}
    \begin{equation}
         P(t,x) = \begin{cases}
            P_L & \frac{x}{t}< D \\
            P_0 & \frac{x}{t}> D 
        \end{cases}, \label{eqn: planar-pressure-solution}
    \end{equation}
    \begin{equation}
         u(t,x) = \begin{cases}
            0 & \frac{x}{t}< D \\
            u_0 & \frac{x}{t}> D 
        \end{cases}. \label{eqn: planar-velocity-solution}
    \end{equation}
\end{subequations}

\subsubsection{General Solution in non-Planar Geometry}

In higher geometries, the initial conditions and the EoS are restricted in order for a solution to exist; see above. If these restrictions are met, the solution is given by: 

\begin{subequations}
    \begin{equation}
        \rho(t,x) = \begin{cases}
            \rho_L & \frac{x}{t}< D \\
           \rho_0 \left(1 - \frac{u_0 t}{x} \right)^m & \frac{x}{t} > D
        \end{cases}, \label{eqn: density-solution}
    \end{equation}
    \begin{equation}
         P(t,x) = \begin{cases}
            P_L & \frac{x}{t}< D \\
            0 & \frac{x}{t}> D 
        \end{cases} ,\label{eqn: pressure-solution}
    \end{equation}
    \begin{equation}
        u(t,x) = \begin{cases}
            0 & \frac{x}{t}< D \\
            u_0 & \frac{x}{t}> D 
        \end{cases}.  \label{eqn: velocity-solution}
    \end{equation}
\end{subequations}

\begin{remark}
    Because of the restrictions, the stiffened gas, Steinberg, and (in general) tabulated EoSs do not admit a solution to the Noh problem in non-planar geometry. Any solution, therefore, derived using these EoSs must take place in planar geometry. We are able derive solutions in non-planar geometry for the Noble-Able and Carnahan-Starling EoSs. 
\end{remark}

\noindent The key observation is the solution is completely determined by \(\rho_L, P_L\) and \(D\). Once these values are known, the solution is known.

\subsubsection{Solving for the Shocked Variables}

\par As is standard in the literature, the \((\rho_L, P_L, D)\) values are found by solving the Rankine-Hugoniot jump conditions. The solution has been found for many equations of state; we refer the reader to \cite{axford_2000, Ramsey2017, Ramsey2018} for examples. We re-derive the analysis for clarity and to highlight its general, implicit form which is agnostic to the particular EoS. Our analysis does not incorporate the restrictions highlighted in section \ref{subsec: general-solution} to simplify the equations, but the reader should still bear in mind that for this approach to me meaningful, the restrictions should still be satisfied when \( m \in \{1,2\}\). 

\par The general Rankine-Hugoniot jump conditions---that is, not specific to the Noh problem---are given by:

\begin{subequations}
    \begin{equation}
       (u_L -D )\rho_L = (u_R - D)\rho_R, \label{eqn: RH-1}
    \end{equation}
    \begin{equation}
        P_L + (u_L- D)\rho_Lu_L = P_R + (u_R - D) \rho_R u_R ,\label{eqn: RH-2}
    \end{equation}
    \begin{equation}
        e_L + \frac{P_L}{\rho_L} + \frac{1}{2}(u_L - D)^2 = e_R + \frac{P_R}{\rho_R} + \frac{1}{2}(u_R- D)^2. \label{eqn: RH-3}
    \end{equation}
\end{subequations}

\noindent We use \eqref{eqn: RH-1} to rewrite \eqref{eqn: RH-2}: 

\begin{equation}
    P_L + (u_L - D)\rho_L u_L = P_R + (u_L - D) \rho_L u_R. \label{eqn: pressure-der-1}
\end{equation}

\noindent Using the assumption \(u_L = 0\) and \(u_R = u_0\), we arrive at the following system of equations: 

\begin{subequations}
    \begin{equation}
        -D \rho_L = (u_0 - D)\rho_R, \label{eqn: pre-Noh-RH-1}
    \end{equation}
    \begin{equation}
        P_L = P_R  - \rho_L D u_0 ,\label{eqn: pre-Noh-RH-2}
    \end{equation}
    \begin{equation}
        e_L + \frac{P_L}{\rho_L} = e_R + \frac{P_R}{\rho_R} + \frac{1}{2} u_0(u_0 - 2D). \label{eqn: pre-Noh-RH-3}
    \end{equation}
\end{subequations}

\noindent To cast the jump conditions in terms of \(\rho_L, P_L, D\), we first substitute \(\rho_R = \rho_0 \left( 1 - \frac{u_0}{D}\right)^m\) into \eqref{eqn: pre-Noh-RH-1}. We keep \eqref{eqn: pre-Noh-RH-2} as is, and using the fact that \eqref{eqn: pre-Noh-RH-2} gives \(D = \frac{P_0 -P_L}{\rho_L u_0}\), we manipulate \eqref{eqn: pre-Noh-RH-3} to obtain the following form of the jump conditions: 

\begin{subequations}
    \begin{equation}
        \rho_L = \rho_0 \left( 1 - \frac{u_0}{D}\right)^{m+1}, \label{eqn: Noh-RH-1}
    \end{equation}
    \begin{equation}
        P_L = P_0 + \rho_L u_0 D, \label{eqn: Noh-RH-2}
    \end{equation}
    \begin{equation}
        e_L + \frac{u_0 P_0}{\rho_L D} = e_R + \frac{1}{2} u_0^2. \label{eqn: Noh-RH-3}
    \end{equation}
\end{subequations}

\noindent Equations \eqref{eqn: Noh-RH-1}-\eqref{eqn: Noh-RH-3} form the system that we propose to solve for the required coefficients. Note that we solve for \(\rho_L, P_L, e_L\), and \(D\), and that the EoS closes the system.

\subsection{Solving with an arbitrary Equation of State}

\par We now describe how to solve \eqref{eqn: Noh-RH-1}-\eqref{eqn: Noh-RH-3} when the equation of state is a black-box. By \say{black-box}, we mean that there exists an object that relates the specific internal energy, density, and pressure. For our purposes, we assume that the relation can be posed as either \(e = e(\rho,P)\) or \(P = P(\rho, e)\). We also assume the black box provides the partial derivatives, though this assumption is not strictly necessary: we can use the function values to approximate the derivatives via finite differences.\footnote{We compared using centered finite differences to exact partial derivatives to solve the jump conditions with the ideal gas, stiffened gas, and Steinberg model for aluminum equations of state. We found that finite differences were mostly able to compute the solution to the same order of error (10E-08 or better) with the same number of Newton iterations (\(\pm 1\) iteration) as using exact partial derivatives. There was one exception with stiffened gas where more iterations were required when using finite differences.} 

\par Under these assumptions, we solve \eqref{eqn: Noh-RH-1}-\eqref{eqn: Noh-RH-3} by employing a Newton iterative solver. Since a Newton solver only requires point-wise values, we only need to assume the EoS is a black-box as described above. If the EoS is posed as \(e = e(\rho, P)\), the residual function is given by: 

\begin{equation}
    \mathbf{F_e}(\rho, P, D) = \begin{pmatrix}
        \rho - \rho_0 \left( 1 - \frac{u_0}{D}\right)^{m+1} \\
        P - P_0 + \rho u_0 D \\
        e(\rho,P) - e_0 - \frac{1}{2} u_0^2 + \frac{u_0 P_0}{\rho D} 
    \end{pmatrix}. \label{form: 3D-energy-residual}
\end{equation}

\noindent The Jacobian matrix of the residual function is then: 

\begin{equation}
    (\mathbf{F'_e})(\rho, P, D) = \begin{pmatrix}
        1 & 0 & -\rho_0(m+1)\left( 1 - \frac{u_0}{D}\right)^m \frac{u_0}{D^2} \\
        u_0D & 1 & \rho u_0 \\
        \frac{\del e}{\del \rho}(\rho, P)  - \frac{u_0 P_0}{\rho^2 D} \; \; \; \; \; & \frac{\del e}{\del P}(\rho, P) & -\frac{u_0 P_0}{\rho D^2} 
    \end{pmatrix}. \label{form: 3D-energy-Jacobian}
\end{equation}

\par If the EoS is posed as \(P = P(\rho, e)\), the residual function becomes: 

\begin{equation}
        \mathbf{F_p}(\rho, e, D) = \begin{pmatrix}
        \rho - \rho_0 \left( 1 - \frac{u_0}{D}\right)^{m+1} \\
        P(\rho,e) - P_0 + \rho u_0 D \\
        e - e_0 - \frac{1}{2} u_0^2 + \frac{u_0 P_0}{\rho D}. \label{form: 3D-pressure-residual}
    \end{pmatrix}.
\end{equation}

\noindent The Jacobian matrix is given by: 

\begin{equation}
   \mathbf{F'_p}(\rho, e,D) = \begin{pmatrix}
        1 & 0 & \; \; \; \; \; -\rho_0(m+1)\left( 1 - \frac{u_0}{D}\right)^m \frac{u_0}{D^2} \\
        \frac{\del P}{\del \rho}(\rho, P) + u_0 D  \; \; \; & \frac{\del P}{\del e}(\rho, e) & \rho u_0 \\
        - \frac{u_0 P_0}{\rho^2 D} & 1 & -\frac{u_0 P_0}{\rho D^2} 
    \end{pmatrix}. \label{form: 3D-pressure-jacobian}
\end{equation}

\noindent Under the additional assumptions \(m =0\) (planar geometry) and \(P_0=0\), the jump conditions can be reduced to a system of two equations: 

\begin{subequations}
       \begin{equation}
        P_L - \rho_0u_0^2 - \rho_0 \frac{P_L}{\rho_L} = 0, \label{eqn: jump-prop-p}
    \end{equation}
        \begin{equation}
        e_L = e_R + \frac{1}{2} u_0^2 = e_0 + \frac{1}{2} u_0^2. \label{eqn: jump-prop-e}
    \end{equation}
\end{subequations}

If we use \(e = e(\rho,P)\) to solve \eqref{eqn: jump-prop-p} - \eqref{eqn: jump-prop-e} for \(\rho_L, P_L\), the residual is given by:

\begin{equation}
    \mathbf{F^s_e}(\rho, P) := \begin{pmatrix}
        P - \rho_0u_0^2 - \rho_0 \frac{P}{\rho}\\
        e(\rho, P) - e_0 - \frac{1}{2}u_0^2
    \end{pmatrix}.  \label{eqn: 2D-energy-residual}
\end{equation}

\noindent The Jacobian is given by: 

\begin{equation}
    \mathbf{(F^{s}_e)'}(\rho,P) := \begin{pmatrix}
    \frac{P}{\rho^2}\rho_0 & 1 - \frac{1}{\rho}\rho_0 \\
    \frac{\del e}{\del \rho}(\rho, P) \; \; \;& \frac{\del e}{\del P}(\rho, P)
    \end{pmatrix}. \label{eqn: 2d-jacobian}
\end{equation}

If we use \(P = P(\rho, e)\), the residual is given by: 

\begin{equation}
    \mathbf{F_p^s}(\rho, e) := \begin{pmatrix}
        P(\rho,e) - \rho_0u_0^2 - \rho_0 \frac{P(\rho, e)}{\rho}\\
        e - e_0 - \frac{1}{2}u_0^2
    \end{pmatrix}.  \label{eqn: 2D-pressure-residual}
\end{equation}

\noindent The Jacobian matrix is given by: 

\begin{equation}
    \mathbf{(F_p^s)'}(\rho,e) := \begin{pmatrix}
        \frac{\del P}{\del \rho}(\rho, e) - \rho_0\left( \frac{\frac{\del P}{\del \rho}(\rho,e) \rho - P(\rho,e)}{\rho^2}\right) \; \; & \frac{\del P}{\del e}(\rho, e) - \rho_0 \frac{\del P}{\del e}(\rho, e) \frac{1}{\rho}\\
        0 & 1
    \end{pmatrix}. \label{eqn: pressure-2d-jacobian}
\end{equation}

\noindent We include these simplified residuals for a practical purpose. Since we are using a Newton solver, we run into the question of choosing an initial guess that is close to the solution. Given the black-box nature of the equation of state, it can be difficult to find a solution if a truly arbitrary set of initial conditions is coupled with a complex equation of state. What these residuals do is allow us to solve a simpler problem: planar Noh with zero initial pressure. We only have to find two variables, and in practice, we can utilize material properties to make an educated guess to their values. However, once we have a solution to the simplified problem, we can use the result to find the solution to the harder problem. For example, we use the result to from the simplified problem as an initial guess for a problem with small initial pressure. This processes is repeated---using the previous result an initial guess for a problem with a larger initial pressure---until the desired initial condition is reached. 
\section{Results} \label{sec: resultgs}
In this section, we present example of semi-analytic solutions derived using the above method. Additionally, we present a verification test to demonstrate the verification capabilities of the semi-analytic solutions. 

\subsection{Examples of solutions}
We demonstrate the ability of our method to derive the semi-analytic solution to the Noh problem with a variety of equations of state. For the sake of brevity, we only present solutions that have yet to be derived. For example, we do not demonstrate the method's ability to solve the Noh problem with ideal gas, as that solution can be found analytically in all geometries \cite{Ramsey2017}. Additionally, we include several solutions for each EoS to demonstrate the method versatility. 

\par We derive solutions for five equations of state: stiffened gas, Noble-Able, Carnahan-Starling, Steinberg, and tabulated aluminum equations of state. Stiffened gas has been analyzed under the assumption that the reference density is equal to the initial density, i.e., \(\rho_\infty = \rho_0\); we extend the analysis by removing this assumption \cite{Boyd_Ramsey_Baty_17, Burnett_17}. Noble-Able and Carnahan-Starling have also been studied, but implicit solutions were derived; see \cite[Equation 54]{Burnett_17} and \cite{Burnett_17}. To the best of our knowledge, no solutions have been derived for Steinberg or tabulated EoSs. We assume that the underlying structure for these EoSs admits a Noh solution; in particular, we demonstrate the the method is capable of deriving solutions for a wide range of complex EoSs. 

\par To compute these solutions, we solve the jump conditions \eqref{eqn: Noh-RH-1}-\eqref{eqn: Noh-RH-3}. We then compute the location of the shock at \(t = 1\)s, and choose a suitable domain to plot the solution. We present both the density and pressure solutions for each equation of state. 

\subsubsection{Stiffened Gas}

\par We begin with stiffened gas, which we consider in the form: 

\begin{equation}
    e(\rho, P) = \frac{P - c_s^2( \rho - \rho_\infty)}{ \rho(\gamma-1)} ~~~ \gamma := \frac{5}{3}. \label{eqn: stiff-gas-eos}
\end{equation}

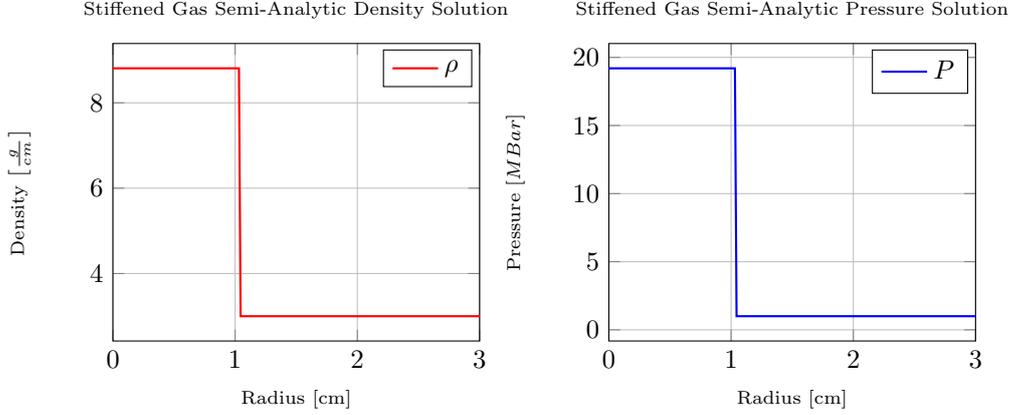
\begin{figure}[H]
    \centering
    \begin{minipage}{0.49\textwidth}
        \centering
        \begin{tikzpicture}
            \begin{axis}[
                width = \textwidth,
                grid = both,
                xmin = 0, xmax = 3,
                xlabel = {Radius [cm]},
                ylabel = {Density \(\left[\frac{g}{cm}\right]\)},
                label style = {font=\footnotesize},
                title = {Stiffened Gas Semi-Analytic Density Solution},
                title style = {font=\footnotesize}
            ]
            \addplot[thick, red] table {data_files/stiff_gas_semi_analytic_density_solution.txt}; 
            \legend{\(\rho\)}
            \end{axis}
        \end{tikzpicture}
    \end{minipage}%
    \hfill
    \begin{minipage}{0.49\textwidth}
        \centering
        \begin{tikzpicture}
            \begin{axis}[
                width = \textwidth,
                grid = both,
                xmin = 0, xmax = 3,
                xlabel = {Radius [cm]},
                ylabel = {Pressure [\(MBar\)]},
                label style = {font=\footnotesize},
                title = {Stiffened Gas Semi-Analytic Pressure Solution},
                title style = {font=\footnotesize}
            ]
            \addplot[thick, blue] table {data_files/stiff_gas_semi_analytic_pressure_solution.txt}; 
            \legend{\(P\)}
            \end{axis}
        \end{tikzpicture}
    \end{minipage}
    \caption{Semi-analytic density and pressure solutions to the planar Noh problem using the stiffened gas equation of state \eqref{eqn: stiff-gas-eos}.}
    \label{fig: stiff-gas-solution}
\end{figure}

\noindent We select \(\rho_\infty= 1\) and use residual \eqref{form: 3D-energy-residual}. Table \ref{tab: stiff_gas_solver_results} displays solutions to the jump conditions for four sets of initial conditions. We also display a complete semi-analytic solution in figure \ref{fig: stiff-gas-solution} for the case with initial conditions \((\rho_0, P_0, u_0) = (3, 1, -2)\). 

\begin{table}[h]
    \centering
    \scriptsize
    \begin{tabular}{cccc|c}
        \(\rho_0\) & \(P_0\) & \(u_0\) & \(m\) & \((\rho_L~ (\text{g/cm}), ~P_L ~(\text{MBar}),~ D ~(\text{cm/s}))\)  \\ \hline
        1 & 0 & -1 & 0 & \((1.8931498239, 2.1196329812, 1.1196329812)\) \\
        3 & 1 & -2 & 0 & \((8.8082886929, 19.1980390272,  1.0330065045)\) \\
        5 & 3 & - 10 & 0 & \((19.5662418895, 674.6297188364,   3.4325943767)\) \\
        10 & 10 & -10 & 0 & \((38.8230631583, 1356.944387732, 3.4694438773)\)  \\
    \end{tabular}
    \caption{\footnotesize EoS: Solutions to the jump conditions using the stiffened gas equation of state \eqref{eqn: stiff-gas-eos}.}
    \label{tab: stiff_gas_solver_results}
\end{table}
\vspace{-10pt}

\subsubsection{Noble-Able}

\par We consider the Noble-Able EoS in the form: 

\begin{equation}
    e(\rho,P) = \frac{P(1 - b\rho)}{\rho(\gamma-1)}, \quad \gamma:= \frac{5}{3}.  \label{eqn: noble-able-eos}
\end{equation}

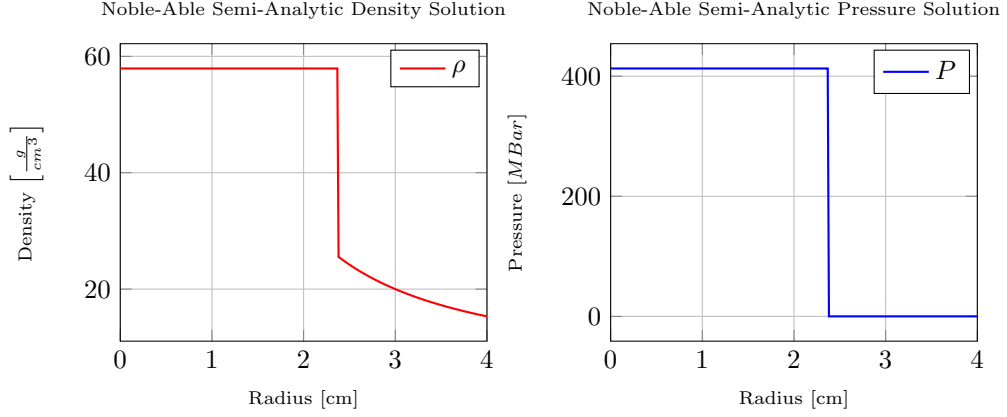
\begin{figure}[H]
    \centering
    \begin{minipage}{0.49\textwidth}
        \centering
        \begin{tikzpicture}
            \begin{axis}[
                width = \textwidth,
                grid = both,
                xmin = 0, xmax = 4,
                xlabel = {Radius [cm]},
                ylabel = {Density \(\left[\frac{g}{cm^3}\right]\)},
                label style = {font=\footnotesize},
                title = {Noble-Able Semi-Analytic Density Solution},
                title style = {font=\footnotesize}
            ]
            \addplot[thick, red] table {data_files/Noble_Able_semi_analytic_density_solution.txt}; 
            \legend{\(\rho\)}
            \end{axis}
        \end{tikzpicture}
    \end{minipage}%
    \hfill
    \begin{minipage}{0.49\textwidth}
        \centering
        \begin{tikzpicture}
            \begin{axis}[
                width = \textwidth,
                grid = both,
                xmin = 0, xmax = 4,
                xlabel = {Radius [cm]},
                ylabel = {Pressure [\(MBar\)]},
                label style = {font=\footnotesize},
                title = {Noble-Able Semi-Analytic Pressure Solution},
                title style = {font=\footnotesize}
            ]
            \addplot[thick, blue] table {data_files/Noble_Able_semi_analytic_pressure_solution.txt}; 
            \legend{\(P\)}
            \end{axis}
        \end{tikzpicture}
    \end{minipage}
    \caption{Semi-analytic density and pressure solutions to the spherical Noh problem using the Noble-Able equation of state \eqref{eqn: noble-able-eos}.}
    \label{fig: noble-able_solution}
\end{figure}

\noindent We select \(b = 0.01\) and used the residual given in \eqref{form: 3D-energy-residual}. Table \ref{tab: noble-able_solver_results} displays the solution to the jump conditions for five sets of initial conditions which range across all three geometries. We also display a complete semi-analytic solution in figure \ref{fig: noble-able_solution} for the spherical case with initial conditions \((\rho_0, P_0, u_0) = (5, 0, -3)\). 

\begin{table}[h]
    \centering
    \scriptsize
    \begin{tabular}{cccc|c}
        \(\rho_0\) & \(P_0\) & \(u_0\) & \(m\) &  (\(\rho_L\) (g/cm\(^{m+1}\)), \(\,P_L\) (MBar), \(\,D\) (cm/s))  \\ \hline
        1 & 0 & -1 & 0 & \((3.8834951456, 1.3468013468, 0.3468013468)\)  \\
        1 & 1 & -1 & 0 & \((1.8808720844, 3.1352386092, 1.1352386092)\)  \\
        1 & 0 & -1 & 1 & \((13.0263855224,  4.992466432 ,  0.3832579977)\)  \\
        1 & 0 & -1 & 2 & \((29.8879859818, 14.2096360138,  0.4754296935)\)  \\
        5 & 0 & -3 & 2 & \((57.9139303375, 412.8249380511, 2.3760831268)\)  
    \end{tabular}
    \caption{\footnotesize EoS: Solutions to the jump conditions using the Noble-Able equation of state \eqref{eqn: noble-able-eos}.}
    \label{tab: noble-able_solver_results}
\end{table}

\subsubsection{Carnahan-Starling}

\par For the Carnahan-Starling EoS, we consider the form: 

\begin{equation}
    e(\rho, P) = \frac{P}{\rho Z(\rho) (\gamma -1)}, \quad Z(\eta) := \frac{1 + \eta + \eta^2 -\eta^3}{(1 - \eta)^3}, \quad \eta(\rho) = b \rho, \quad \gamma := \frac{5}{3}. \label{eqn: carnahan-starling-eos}
\end{equation}

\begin{figure}[H]
    \centering
    \begin{minipage}{0.49\textwidth}
        \centering
        \begin{tikzpicture}
            \begin{axis}[
                width = \textwidth,
                grid = both,
                xmin = 0, xmax = 2,
                xlabel = {Radius [cm]},
                ylabel = {Density \(\left[\frac{g}{cm^2}\right]\)},
                label style = {font=\footnotesize},
                title = {Carnahan-Starling Semi-Analytic Density Solution},
                title style = {font=\footnotesize}
            ]
            \addplot[thick, red] table {data_files/Carnahan_Starling_semi_analytic_density_solution.txt}; 
            \legend{\(\rho\)}
            \end{axis}
        \end{tikzpicture}
    \end{minipage}%
    \hfill
    \begin{minipage}{0.49\textwidth}
        \centering
        \begin{tikzpicture}
            \begin{axis}[
                width = \textwidth,
                grid = both,
                xmin = 0, xmax = 2,
                xlabel = {Radius [cm]},
                ylabel = {Pressure [\(MBar\)]},
                label style = {font=\footnotesize},
                title = {Carnahan-Starling Semi-Analytic Pressure Solution},
                title style = {font=\footnotesize}
            ]
            \addplot[thick, blue] table {data_files/Carnahan_Starling_semi_analytic_pressure_solution.txt}; 
            \legend{\(P\)}
            \end{axis}
        \end{tikzpicture}
    \end{minipage}
    \caption{Semi-analytic density and pressure solutions to the cylindrical Noh problem using the Carnahan-Starling equation of state \eqref{eqn: carnahan-starling-eos}.}
    \label{fig: carnahan-starling_solution}
\end{figure}
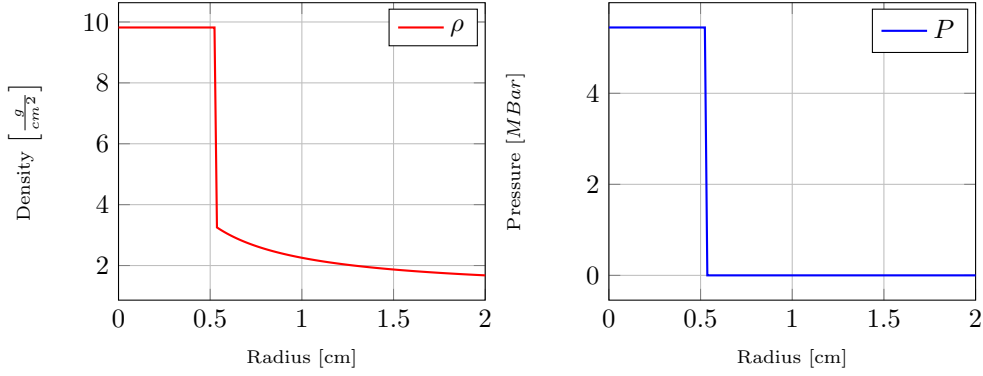

\noindent We use \(b = 0.01\) and the residual given in \eqref{form: 3D-energy-residual}. Table \ref{tab: carnahan-starling_solver_results} displays the solution to the jump conditions for four sets of initial conditions ranging across two geometries. We also display a complete semi-analytic solution in figure \ref{fig: carnahan-starling_solution} for the cylindrical case with initial conditions \((\rho_0, P_0, u_0) = (1.1, 0, -1.05)\). 

\begin{table}[h]
    \centering
    \scriptsize
    \begin{tabular}{cccc|c}
        \(\rho_0\) & \(P_0\) & \(u_0\) & \(m\) & (\(\rho_L\) (g/cm\(^{m+1}\)), \(\,P_L\) (MBar), \(\,D\) (cm/s))  \\ \hline
        1 & 0 & -1 & 0 & \((3.5918818886, 1.3858200501, 0.3858200501)\)  \\
        1.5 & 0.2 & -1 & 0 & \((4.1633168433, 2.5448112382, 0.5632074921)\) \\
        1 & 0 & -1 & 1 & \((9.2359068339, 4.5294847 , 0.4904212203)\) \\
        1.1 & 0 & -1.05 & 1 & \((9.8192441455, 5.4462459744, 0.5282383095)\) 
    \end{tabular}
    \caption{\footnotesize EoS: Solutions to the jump conditions using the Carnahan-Starling equation of state \eqref{eqn: carnahan-starling-eos}.}
    \label{tab: carnahan-starling_solver_results}
\end{table} 

\subsubsection{Steinberg}
\par The Steinberg (sometimes called Mie-Gruneisen or simply Gruneisen) equations of state can be found in \cite{steinberg96}. These equations take the following \(P = P(\rho,e)\) form:

\begin{equation}
    P(\rho,e) = P_H(\rho) + \rho \Gamma(\rho) ( e - e_H(\rho)), \label{eqn: steinberg-eos}
\end{equation}

\noindent with: 

\begin{multline}
    \eta(\rho) = 1 - \frac{\rho^\ast}{\rho}, \quad \Gamma(\rho) = \begin{cases}
        \Gamma^\ast & \eta \leq 0 \\ \Gamma_0( 1- \eta) + b \eta & 0 \leq \eta <1 
    \end{cases}, \quad e_H(\rho) = \begin{cases}
        0 & \rho < \rho^\ast\\
        \frac{\eta(P_H(\rho) + P^\ast)}{2 \rho^\ast} & 0 \leq \eta < 1 
    \end{cases}, \\
    \hspace*{\fill} P_H(\rho) = \rho^\ast + c_0^2 \eta \begin{cases}
        \rho & \rho < \rho^\ast \\
        \frac{\rho^\ast}{(1 - s_1 \eta - s_2 \eta^2 - s_3 \eta^3)^2} & \rho \geq \rho^\ast
    \end{cases}. \hspace*{\fill}
\end{multline}

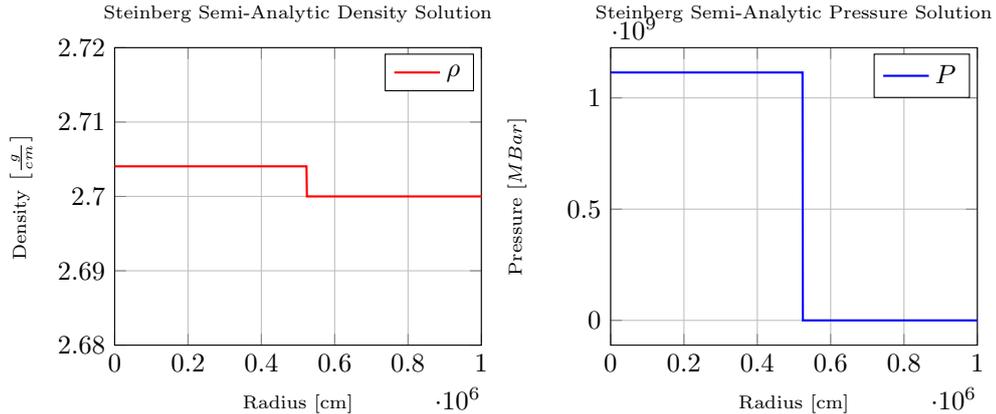
\begin{figure}[H]
    \centering
    \begin{minipage}{0.49\textwidth}
        \centering
        \begin{tikzpicture}
            \begin{axis}[
                width = \textwidth,
                grid = both,
                xmin = 0, xmax = 1000000,
                ymin = 2.68, ymax = 2.72,
                xlabel = {Radius [cm]},
                ylabel = {Density \(\left[\frac{g}{cm}\right]\)},
                label style = {font=\footnotesize},
                title = {Steinberg Semi-Analytic Density Solution},
                title style = {font=\footnotesize}
            ]
            \addplot[thick, red] table {data_files/Steinberg_semi_analytic_density_solution.txt}; 
            \legend{\(\rho\)}
            \end{axis}
        \end{tikzpicture}
    \end{minipage}%
    \hfill
    \begin{minipage}{0.49\textwidth}
        \centering
        \begin{tikzpicture}
            \begin{axis}[
                width = \textwidth,
                grid = both,
                xmin = 0, xmax = 1000000,
                xlabel = {Radius [cm]},
                ylabel = {Pressure [\(MBar\)]},
                label style = {font=\footnotesize},
                title = {Steinberg Semi-Analytic Pressure Solution},
                title style = {font=\footnotesize}
            ]
            \addplot[thick, blue] table {data_files/Steinberg_semi_analytic_pressure_solution.txt}; 
            \legend{\(P\)}
            \end{axis}
        \end{tikzpicture}
    \end{minipage}
    \caption{Semi-analytic density and pressure solutions to the planar Noh problem using the Steinberg equation of state for aluminum \eqref{eqn: steinberg-eos}.}
    \label{fig: steinberg_solution}
\end{figure}

\noindent We test using the aluminum (6061-T6) material found in \cite{steinberg96}. The parameters for this material are: \(\rho^\ast = 2.703, ~P^\ast = 0, ~\Gamma^\ast = 1.97, ~b = 0.48, ~c_0 = 0.524\times 10^6, ~s_1 = 1.4, ~s_2 = s_3 = 0\). Table \ref{tab: steinberg_solver_results} displays solutions to the jump conditions for two initial conditions in planar geometry. We also display the complete semi-analytic solution in figure \ref{fig: steinberg_solution} for the case with initial conditions \((\rho_0, P_0, u_0) = (2.7, 20, -1.5\times 0.524\times10^3)\). 

\begin{table}[h]
    \centering
    \scriptsize
    \begin{tabular}{cccc|ccccc}
        \(\rho_0\) & \(P_0\) & \(u_0\) & \(m\) & (\(\rho_L\) (g/cm), \(\,e_L\)  (cm\(^2\)/s\(^2\)), \(\,D\) (cm/s)) \\ \hline
        2.7 & 0 & \(-1.5\times 0.524\times10^3\) & 0 & \((2.7040480683, 155174098.23, 524250.65318)\) \\
        2.7 & 20 & \(-1.5\times 0.524\times10^3\) & 0 & \((2.7040480683, 15517410.200, 52425.005320)\) \\
    \end{tabular}
    \caption{\footnotesize EoS: Solutions to the jump conditions using the Steinberg equation of state \eqref{eqn: steinberg-eos} and pressure residual \eqref{form: 3D-pressure-residual}.}
    \label{tab: steinberg_solver_results}
\end{table}
\vspace{-15pt}

\subsubsection{Aluminum Tabulated EoS}

\par
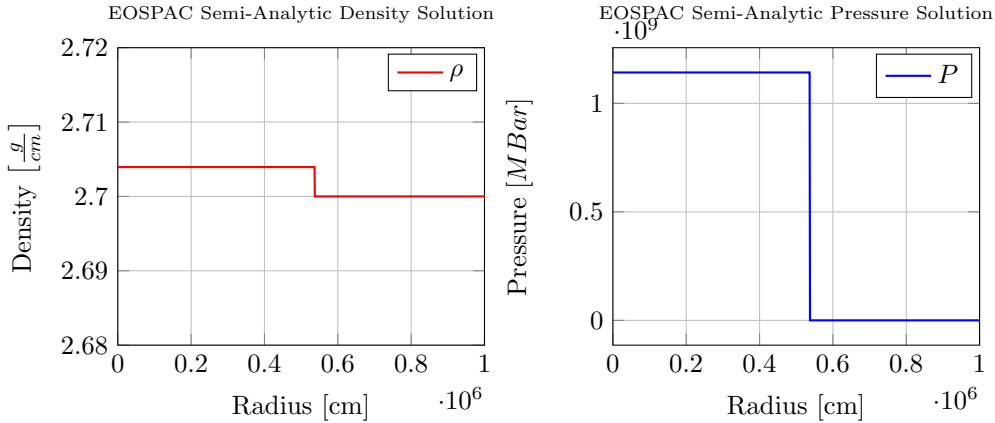
\begin{figure}[H]
    \centering
    \begin{minipage}{0.49\textwidth}
        \centering
        \begin{tikzpicture}
            \begin{axis}[
                width = \textwidth,
                grid = both,
                xmin = 0, xmax = 1000000,
                ymin = 2.68, ymax = 2.72,
                xlabel = {Radius [cm]},
                ylabel = {Density \(\left[\frac{g}{cm}\right]\)},
                title = {EOSPAC Semi-Analytic Density Solution},
                title style = {font=\footnotesize}
            ]
            \addplot[thick, red] table {data_files/EOSPAC_semi_analytic_density_solution.txt};
            \legend{\(\rho\)}
            \end{axis}
        \end{tikzpicture}
    \end{minipage}%
    \hfill
    \begin{minipage}{0.49\textwidth}
        \centering
        \begin{tikzpicture}
            \begin{axis}[
                width = \textwidth,
                grid = both,
                xmin = 0, xmax = 1000000,
                xlabel = {Radius [cm]},
                ylabel = {Pressure [\(MBar\)]},
                title = {EOSPAC Semi-Analytic Pressure Solution},
                title style = {font=\footnotesize}
            ]
            \addplot[thick, blue] table {data_files/EOSPAC_semi_analytic_pressure_solution.txt}; 
            \legend{\(P\)}
            \end{axis}
        \end{tikzpicture}
    \end{minipage}
    \caption{Semi-analytic density and pressure solutions to the planar Noh problem using the EOSPAC equation of state for aluminum (3720).}
    \label{fig: eospac_solution}
\end{figure}

 Our final EoS is a tabulated equations of state. Tabulated EoSs do not have a closed equation form; instead, they are modeled using tables of values and interpolation techniques. We use the table for aluminum, \verb|SESAME| number 3720, and use the \verb|singularity-eos| equation of state library \cite{sing-eos} that incorporates the \verb|EOSPAC| library. To find a solution, we use the residual given by \eqref{form: 3D-pressure-residual}. Table \ref{tab: eospac_solver_results} displays solutions to the jump conditions for two initial conditions in planar geometry. We also display the complete semi-analytic solution in figure \ref{fig: eospac_solution} for the case with initial conditions \((\rho_0, P_0, u_0) = (2.7, 20, -1.5\times 0.524\times10^3)\).

\begin{table}[h]
    \centering
    \scriptsize
    \begin{tabular}{cccc|ccccc}
        \(\rho_0\) & \(P_0\) & \(u_0\) & \(m\) & (\(\rho_L\) (g/cm), \(\,e_L\) (cm\(^2\)/s\(^2\)), \(\,D\) (cm/s)) \\ \hline
        2.7 & 0 & -786 & 0 & \((2.7039485211, 290525.89791, 537467.05202)\) \\
        2.7 & 15 & -786 & 0 & \((2.7039485211, 290528.54944, 537467.05203\) \\
    \end{tabular}
    \caption{\footnotesize EoS: Solutions to the jump conditions using the aluminum EOSPAC tabulated equation of state and pressure residual \eqref{form: 3D-pressure-residual}.}
    \label{tab: eospac_solver_results}
\end{table}
\vspace{-15pt}

\begin{remark}[Using the \(e = e(\rho, P)\) vs the \(P = P(\rho, e)\) residual.]
    To compare the solver's performance based on the different residuals, we attempt to find solutions for Carnahan-Starling using both the energy-based residual \eqref{form: 3D-energy-residual} and pressure-based residual \eqref{form: 3D-pressure-residual}. The pressure-based converges with fewer iterations and is able to solve more problems. 
    \par For example, with initial conditions \(\rho_0 = 1.1, ~P_0 = 0, ~u_0 = -1.05\) in cylindrical geometry (\(m =1\)) and an initial guess of \((\rho, P, D) = (9.9, 5.44, 0.53)\), the energy-based Newton solver took 136 iterations to reach the solution: \((\rho_L, P_L, D) = (9.8192441455, 5.4462459744, 0.5282383095)\). Using the same initial conditions with an initial guess of \((\rho, e, D) = (14, 1.8, 2.8)\), the pressure-based Newton solver converged to the solution \((\rho_L, e_L, D) = (9.8192441455, 0.55125, 0.5282383095)\) in 10 iterations.
    \par When we increase the initial velocity to \(u_0 = -1.06\) and use the same initial guess, the energy-solver does not converge with three-thousand iterations. On the other hand, the pressure-based solver converges to a solution in 10 iterations. 
    \par We suspect this behavior stems from the structures of the different formulations. To simplify our analysis, let us examine the simplified residuals. Plotting the second equation (with constant pressure) of the simplified energy residual \eqref{eqn: 2D-energy-residual} shows that, regardless of the initial conditions, the roots exist close to the asymptotes. Additionally, the derivative (\( \frac{\del e}{\del \rho}\)) increases dramatically: it is near zero, then quickly jumps as it approaches the asymptotes. As such, the area of convergence, from a theoretical perspective, can be quite small. On the other hand, plotting the second equation of the simplified pressure residual \eqref{eqn: 2D-pressure-residual} reveals that the roots are comparatively far from the asymptote. Also, the derivative (\(\frac{\del P}{\del \rho}\)) is much more well behaved, i.e., is bounded on a comparatively larger set around the root. As such, the area of convergence can be much larger. Of course, this is only a heuristic analysis; more work should be done to better understand what residual is better suited for a particular EoS.
\end{remark}

\subsection{Verification Example}

\par In this section, we perform a convergence test using a semi-analytic solution derived using the above method. For this test, we solve the Noh problem with the stiffened gas equation of state \eqref{eqn: stiff-gas-eos} in planar geometry. The EoS parameters are \(\rho_\infty =1\), \(c_s = \sqrt{\gamma}\), and \(\gamma = 5/3\) with initial conditions \((\rho_0, P_0, u_0) = (2, 1, -1)\).  We note that the stiffened gas solutions have been analyzed in the past \cite{Ramsey2017, axford_2000, Burnett_17}; however, explicit solutions assumed \(\rho_0 = \rho_\infty\). 

\par
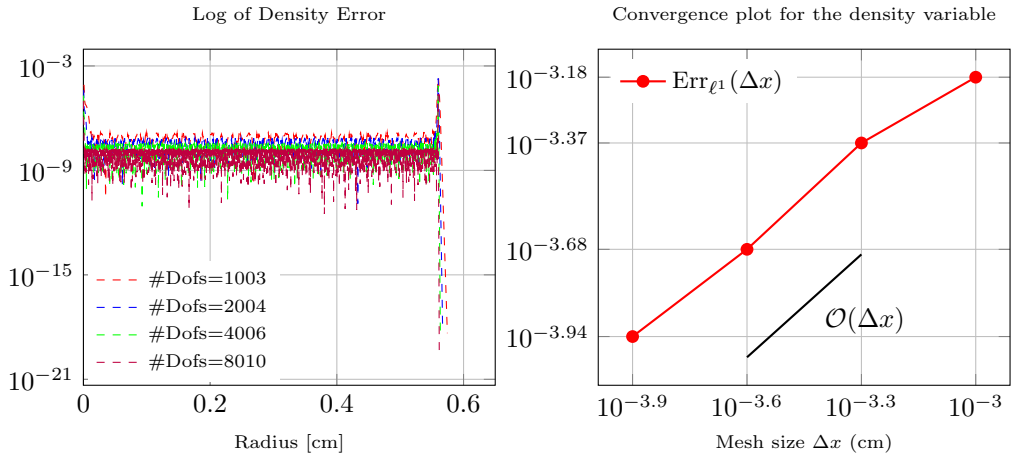
\begin{figure}[H]
\centering
\begin{minipage}[t]{0.49\textwidth}
\centering
\begin{tikzpicture}
\begin{axis}[
    width=7cm,
    grid=both,
ymode = log, xmin = 0, xmax=0.65,
    xlabel={Radius [cm]},
    label style ={font=\footnotesize}, 
    title = {\footnotesize Log of Density Error},
    legend style={
        at={(0.01,0.01)},    
        anchor=south west,
        font=\footnotesize,
        fill=white,
        draw=none
    },
    clip=false             
]
            \addplot[dashed, thin, red] table {data_files/density_errors_1003.txt}; 
            \addplot[dashed, thin, blue] table {data_files/density_errors_2004.txt}; 
            \addplot[dashed, thin, green] table {data_files/density_errors_4006.txt}; 
            \addplot[dashed, thin, purple] table {data_files/density_errors_8010.txt}; 

\legend{
    \#Dofs=1003,
    \#Dofs=2004,
    \#Dofs=4006,
    \#Dofs=8010
}
\end{axis}
\end{tikzpicture}
\end{minipage}
\hfill
\begin{minipage}[t]{0.49\textwidth}
\centering
\begin{tikzpicture}
\begin{loglogaxis}[
    width=7cm,
    xlabel={Mesh size $\Delta x$ (cm)},
    label style = {font=\footnotesize},
    title={\footnotesize Convergence plot for the density variable}, 
    grid=both,
    legend pos=north west,
    legend style={
        font=\scriptsize,
        fill=white,
        draw=none
    },
    clip=false,
    xtick={1/1003, 1/2004, 1/4006, 1/8010},
    ytick={0.000659856, 0.000423641, 0.000206939, 0.000114984}
]
\addplot[
    color=red,
    mark=*,
    thick
] coordinates {
    (1/1003, 0.0006598563578091181)
    (1/2004, 0.0004236409960581233)
    (1/4006, 0.00020693898187613497)
    (1/8010, 0.00011498379257579585)
};
\addlegendentry{$\text{Err}_{\ell^1}(\Delta x)$}

\addplot[
    color=black,
    thick
] coordinates {
    (1/2004, 2e-4)
    (1/4006, 1e-4)
};
\node at (axis cs:1/2650,1.1e-4) [anchor=south west] {$\mathcal{O}(\Delta x)$};
\end{loglogaxis}
\end{tikzpicture}
\end{minipage}
\caption{Numerical results for the density variable \(\rho\) for the Noh problem using the stiffened gas equation of state \eqref{eqn: stiff-gas-eos}. Left: Pointwise relative error for the density variable on four meshes. Right: convergence plot for the discrete \(L^1\)-error, \(\text{Err}_{\ell^1}\)}
\label{fig:stiff_gas_density_verification}
\end{figure}

 We solve this problem numerically using the \verb|xRAGE| Eulerian hydrodynamics code \cite{Gittings_2008}. We set the final time to be \(T = 0.5\). For our solver, we use the pressure-based residual \eqref{form: 3D-pressure-residual}. The resulting values are: \(\rho_L = 3.786299647843569\), \(P_L = 5.239265962364613\), and \(D =  1.119632981182307\). We run the simulation on four uniform meshes and compute the error for the density and pressure variables. The mesh size \(\Delta x\) is given by the quantity \(\frac{1}{\#\text{Dofs}}\). For a given mesh with associated mesh size \(\Delta x\), we compute the error using the discrete \(L^1\)-norm: \(\| u_h \|_{\ell^1(\Delta x)} = \sum_{i = 0}^N \vert u_i \vert\), where \(u_i\) denotes the value of \(u_h\) at the \(i\)-th degree of freedom. The numerical error is computed via the quantity \(\text{Err}_{\ell^1}(\Delta x) = \| u_{\Delta x} - u_s\|_{\ell^1} / \| u_s\|_{\ell^1}.\) Here, \(u_{\Delta x}\) denotes the computed numerical solution using a mesh size \(\Delta x \) and \(u_s\) the corresponding semi-analytic solution. The results for the density variable are displayed in figure \ref{fig:stiff_gas_density_verification}. The left figure shows the log of the point-wise relative error given by each mesh; we have limited the domain to \([0, 0.65]\) since the error is zero outside this region. The right figure displays the convergence plot with a reference first order slope \(\mathcal{O}(\Delta x)\).

\par
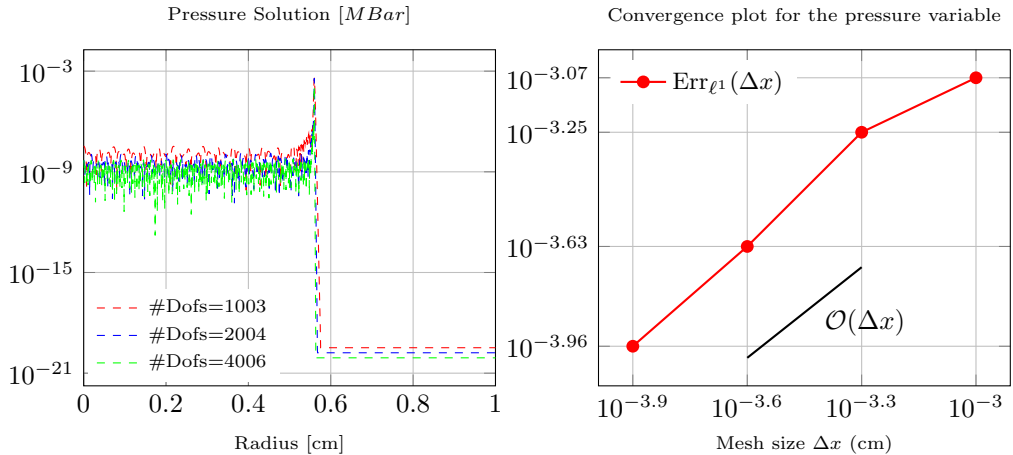
\begin{figure}[H]
\centering
\begin{minipage}[t]{0.49\textwidth}
\centering
\begin{tikzpicture}
\begin{axis}[
    width=7cm,
    grid=both,
    xmin=0, xmax=1.0,
 ymode = log,
    xlabel={Radius [cm]},
    label style={font=\footnotesize}, 
    title = {\footnotesize Pressure Solution \([MBar]\)},
    legend style={
        at={(0.01,0.01)},    
        anchor=south west,
        font=\footnotesize,
        fill=white,
        draw=none
    },
    clip=false             
]

            \addplot[dashed, thin, red] table {data_files/pressure_errors_1003.txt}; 
            \addplot[dashed, thin, blue] table {data_files/pressure_errors_2004.txt}; 
            \addplot[dashed, thin, green] table {data_files/pressure_errors_4006.txt}; 

\legend{
    \#Dofs=1003,
    \#Dofs=2004,
    \#Dofs=4006,
}
\end{axis}
\end{tikzpicture}
\end{minipage}
\hfill
\begin{minipage}[t]{0.49\textwidth}
\centering
\begin{tikzpicture}
\begin{loglogaxis}[
    width=7cm,
    xlabel={Mesh size $\Delta x$ (cm)},
    label style ={font=\footnotesize}, 
    title={\footnotesize Convergence plot for the pressure variable},
    grid=both,
    legend pos=north west,
    legend style={
        font=\scriptsize,
        fill=white,
        draw=none
    },
    clip=false,
    xtick={1/1003, 1/2004, 1/4006, 1/8010},
    ytick={0.0008510822179301525, 0.0005611030962027185, 0.00023434888599370722, 0.00010927231803254066}
]
\addplot[
    color=red,
    mark=*,
    thick
] coordinates {
    (1/1003, 0.0008510822179301525)
    (1/2004, 0.0005611030962027185)
    (1/4006, 0.00023434888599370722)
    (1/8010, 0.00010927231803254066)
};
\addlegendentry{$\text{Err}_{\ell^1}(\Delta x)$}

\addplot[
    color=black,
    thick
] coordinates {
    (1/2004, 2e-4)
    (1/4006, 1e-4)
};
\node at (axis cs:1/2650,1.1e-4) [anchor=south west] {$\mathcal{O}(\Delta x)$};
\end{loglogaxis}
\end{tikzpicture}
\end{minipage}
\caption{Numerical results for the pressure variable \(P\) for the Noh problem using the stiffened gas equation of state \eqref{eqn: stiff-gas-eos}. Left: pointwise relative error for the pressure variable on three meshes. Right: convergence plot for the discrete \(L^1\)-error, \(\text{Err}_{\ell^1}\).}
\label{fig:stiff_gas_pressure_verification}
\end{figure}

 The results for the pressure variable are displayed in figure \ref{fig:stiff_gas_pressure_verification}. The left figure shows the log of the point-wise relative error given by the three meshes. The right figure displays the convergence plot with a reference first order slope \(\mathcal{O}(\Delta x)\). 

\par These results demonstrate the near first-order accuracy in the \(L^1\)-norm, as expected of finite volume methods for problems featuring shocks \cite{Motheau_20, Siklosi_05}. Furthermore, these results illustrate how the semi-analytic solutions derived using the above method may be used to conduct verification tests.

\section{Conclusion} \label{sec: conclusion}

This paper introduces a method for deriving semi-analytic solutions to the Noh problem with a black-box equation of state. The key to the algorithm is to recognize that the solution is completely determined by the Rankine-Hugoniot jump conditions. Solving that system can be done with a Newton solver, and since the solver only requires pointwise values of the equation of state, a black-box EoS can be assumed. We include the theory to describe the method. New semi-analytic solutions to the Noh problem have been derived for a wide range of equations of state, including tabulated EoSs. In addition, a convergence test was performed using using a semi-analytic solution for the stiffened gas equation of state. Experiments also indicate that the pressure-based residual is better at solving the jump-conditions. Future work includes extending the method to more problems such as the Sedov blast problem. Another avenue includes generalizing the Noh problem to alleviate the constraints on the equation of state in higher dimensions. Additional work on this particular problem could include nondimensionalizing the residuals. Such work could help improve solvers for Steinberg or tabulated problems where the variables exist on drastically different scales (for example, density is \(\mathcal{O}(1.0^{1})\) but the pressure is \(\mathcal{O}(1.0^{10})\) in our EOSPAC result \ref{fig: eospac_solution}).

\section*{Acknowledgements}
This work was performed under the auspices of the United State Department of Energy by Los Alamos National Security, LLC.

\end{document}